# Combining advanced photoelectron spectroscopy approaches to analyse deeply buried GaP(As)/Si(100) interfaces: Interfacial chemical states and complete band energy diagrams

O. Romanyuk[1,*], A. Paszuk[2], I. Gordeev[1], R.G. Wilks[3,4], S. Ueda[5,6,7], C. Hartmann[3], R. Félix[3], M. Bär[3,4,8,9], C. Schlueter[10], A. Gloskovskii[10], I. Bartoš[1], M. Nandy[2], J. Houdková, P. Jiříček[1], W. Jaegermann[11], J.P. Hofmann[11], T. Hannappel[2]

[1]*FZU – Institute of Physics of the Czech Academy of Sciences, Prague, Czech Republic*
[2]*Dep: Fundamentals of Energy Materials, Institute of Physics, Ilmenau University of Technology, Ilmenau, Germany*
[3]*Department Interface Design, Helmholtz-Zentrum Berlin für Materialien und Energie GmbH, Berlin, Germany*
[4]*Energy Materials In-Situ Laboratory Berlin (EMIL), Helmholtz-Zentrum Berlin für Materialien und Energie GmbH, Berlin, Germany*
[5]*Synchrotron X-ray Station at SPring-8, National Institute for Materials Science (NIMS), Hyogo, Japan*
[6]*Research Center for Functional Materials, NIMS, Tsukuba, Japan*
[7]*Research Center for Advanced Measurement and Characterization, NIMS, Tsukuba, Japan*
[8]*Department of Chemistry and Pharmacy, Friedrich-Alexander-Universität Erlangen-Nürnberg, Erlangen, Germany*
[9]*Helmholtz-Institute Erlangen-Nürnberg for Renewable Energy (HI ERN), Berlin, Germany*
[10]*Deutsches Elektronen-Synchrotron DESY, Ein Forschungszentrum der Helmholtz-Gemeinschaft, Hamburg, Germany*
[11]*Surface Science Laboratory, Department of Materials and Earth Sciences, Technical University of Darmstadt, Darmstadt, Germany*

*corresponding author, e-mail: romanyuk@fzu.cz*

**Abstract**

The epitaxial growth of the polar GaP(100) on the nonpolar Si(100) substrate suffers from inevitable defects at the antiphase domain boundaries, resulting from mono-atomic steps on the Si(100) surface. Stabilization of Si(100) substrate surfaces with arsenic is a promising technological step enabling the preparation of Si substrates with double atomic steps and reduced density of the APDs. In this paper, 4 – 50-nm-thick GaP epitaxial films were grown on As-terminated Si(100) substrates with different types of doping, miscuts, and As-surface termination by metalorganic vapor phase epitaxy. The GaP(As)/Si(100) heterostructures were investigated by X-ray photoelectron spectroscopy (XPS) combined with gas cluster ion beam (GCIB) sputtering and by hard X-ray photoelectron spectroscopy (HAXPES). We found residuals of arsenic atoms in the GaP lattice (~0.2-0.3 at.%) and a localization of As atoms at the GaP(As)/Si(100) interface (~1 at.%). Deconvolution of core level peaks revealed interface core level shifts. In As core levels, chemical shifts between 0.5-0.8 eV were measured and identified by angle-resolved XPS measurements. Similar valence band offset (VBO) values of 0.6 eV were obtained, regardless of the doping type of Si substrate, Si substrate miscut or type of As-terminated Si substrate surface. The band alignment diagram of the heterostructure was deduced.

# 1. Introduction

Epitaxial integration of III-V semiconductor materials with silicon opens up attractive prospects in the production of inexpensive and highly efficient optoelectronic devices. To overcome the lattice mismatch between III-V materials and Si substrates, a thin, optically transparent buffer layer of GaP can be conventionally grown. Although the mismatch between the polar GaP and non-polar Si crystal lattices is small (0.37%), defects such as antiphase domain (APD) boundaries arise at the interface caused by the difference in their polarities [1]. A general approach for the prevention of defects at the GaP/Si(100) density is to prepare a dimerized, single-domain Si(100) substrate with double-layer-steps on the surface prior to GaP growth [2]. Afterwards, single-domain GaP/Si templates allow a graded pathway to the deposition of other III-V semiconductor layer structures with tunable bandgaps [3,4]. The application of such 'virtual' III-V substrates reduces the density of defects in epitaxial III-V films and increases the performance of (opto-)electronic devices such as solar cells and water splitting photoelectrochemical cells [4,5]. In addition, virtual single-domain GaP/Si templates [6] could potentially reduce the cost of ferromagnetic spintronic microdevices grown on expensive GaP(100) bulk substrates [7,8].

Metalorganic vapor phase epitaxy (MOVPE) is an industry-relevant deposition technique that provides well-defined, high-purity epitaxy. The clean Si(100) surface can be prepared either in a "clean" hydrogen environment (without III-V precursor residuals in the MOVPE reactor) or in a reactor with As precursor residuals. The preparation of Si substrates in As ambience is technologically preferred because As is generally present when preparing III-V layer structures [6]. In addition, annealing the Si surface in an As-rich MOVPE reactors allows to reduce the maximum temperature required for Si deoxidation and accelerates the preparation of the well-defined substrate surface compared to the preparation in a pure hydrogen environment [6,9]. Preliminary deposition of As on the non-polar Si substrate also supports monitoring and controlling the polarity of the GaP heterointerface [10].

As-terminated Si(100) surfaces are prepared by annealing of the Si(100) substrates in tertiarybutylarsine (TBAs) and/or background partial pressure of As (As precursor residuals from the reactor walls) at elevated temperatures [11]. Specific annealing procedures lead to the formation of either Si(100):As (1×2) (A-type) or (2x1) (B-type) surface reconstructions [12].

Two As adsorption modes were observed on the Si surface in ultra-high vacuum (UHV). Additive adsorption of As onto the top layer of the Si dimers switches the orientation of the surface domain from A-type (1x2) to B-type (2x1) [12], in contrast to the adsorption of As in the replacive growth mode, where Si dimers are replaced by As dimers on top and no change in the orientation of the surface domain occurs [13]. The dependence of the Si(100):As surface domain orientation on specific MOVPE preparation procedures was also observed in H- and As-rich ambience [12].

In the past, the atomic structure of the Si(100):As-(2x1) surface was investigated in detail. It has been suggested that the Si surface is terminated by symmetric As-As dimers at the top of the Si lattice [14,15]. Surface structures consisting of Si dimers on the underlying As atoms and As-Si heterodimers were also considered [16,17]. Scanning tunneling microscopy (STM) pattern simulations revealed asymmetric protrusions on Si(100)-(1x2) and zigzag protrusions on c(4x2) surfaces terminated with mixed As-Si dimers. Thus, the inclusion of As in the surface reconstruction and fine-tuning of the As binding configuration at the surface are key parameters to advanced GaP nucleation conditions.

Photoelectron spectroscopic studies of Si(100):As-(2x1) surfaces revealed a surface core level shift (0.4 eV) in both Si 2p and As 3d core level peaks [18,19]. In the present paper, information depth of photoelectron spectroscopy was increased and varied in order to investigate the chemical shifts at buried GaP(As)/Si interfaces.

Depending on Si substrate miscut orientations, the type of As source (direct TBAs precursor supply vs. arsenic back pressure), and annealing temperatures, the ratio of the majority to the minority domain on the Si(100):As surface can be controlled. In particular, the A-type (1x2) surface has been precisely prepared resulting in homogeneously broad, atomically flat, two-layer stepped terraces on Si(100) substrates with miscut orientations of 0.1° and 4° towards the [011] direction [12]. The majority of domains on the Si(100):As surface subsequently determine the domain orientation of the adjacent GaP(As) lattice. Therefore, the growth of GaP on the surface of Si(100):As (1x2) A-type results in the well-known GaP(100)-(1x2) surface reconstruction with H-passivated buckled P-dimers on top [20] and the layer sequence Ga-P-…-Ga-As-Si-Si-…(Si-bulk) at the interface below the buckled P-dimers. The orientation of surface dimers before and during deposition is consistently controlled by reflection anisotropy spectroscopy (RAS) in the MOVPE reactor and a switch of the RAS peak sign clearly shows the reverse domain orientation of the top layer relative to the Si substrate. Conversely, the same domain orientation on the surface of the top layer and the substrate leads to the same sign in the RA spectra [6].

Atomic structure, interface polarity, defect density, and electronic structure at the critical interfaces are particularly important for the functionality of the complete device. Charge carrier transport can be either negatively influenced: charge carriers can recombine non-radiatively as they pass through a heterojunction with an inappropriate electronic level alignment or defects (interfacial) electronic states being present in the band gap. Charge localization at the interface [21,22] and formation of interfacial electronic states [23–26] could especially occur at abrupt and well-ordered heterovalent interfaces. The existence of internal electronic states at the interface determines also the charge carriers dynamics at a heterojunction and could, for example, open conductive channels at the GaP/Si(100) interface [26].

In the past, a wide energetic range (0.3 – 1.1 eV) of valence band offset (VBO) values have been determined for GaP/Si heterostructures [27]. The diversity of VBOs may be related to a difference in the atomic structure of the interface (interface polarity, abruptness, defect density, reconstruction, etc.) [23,28] resulting in a variation of the electronic structure and band bending at the interface. Particularly strong band bending was observed at the abrupt single-domain GaP/i-Si(100) interfaces (with low defect density), while the band bending was lower at the two-domain GaP/n-Si(100) interfaces (with higher defect density) [29].

Charge compensation and barrier reduction at the interface can also be induced by mixing atoms at the III-V / Si interface, where the anion-Si atomic pairs coexist with the cation-Si atomic pairs in the thin interfacial layer [21,23,30]. Arsenic on the surface of the Si substrate predetermines the formation of the As-Si (anion-Si) atomic pair and reduces the formation of the Ga-Si (cation-Si) pairs at the interface.

In this paper, a combination of destructive and non-destructive photoelectron spectroscopy-based depth profiling methods was used to obtain information about the chemical composition, electronic states and the band alignment at the buried GaP(As)/Si interfaces. X-ray photoelectron spectroscopy (XPS) and hard X-ray photoelectron spectroscopy (HAXPES) studies have been performed in the past on As-free, single- and two-domain GaP/Si(100) heterostructures [29,31,32], where interface-related chemical shifts were observed in P 2p and Si 2p core levels. Interdiffusion of P into the i-Si buffer layer (100) was suggested based on HAXPES and secondary ion microscopy (SIMS) measurements. Here, we extend our study of single-domain GaP(As)/Si(100) heterostructures prepared in As-rich environment on commercial Si (100) wafers with different doping types and miscut orientations.

## 2. Experimental details
## 2.1 Sample preparation

All samples were grown in a horizontal MOVPE reactor (Aixtron AIX-200) with $H_2$ as process carrier gas. We employed vicinal Si(100) substrates with 0.1° and 4° miscut orientations towards the [011] direction. The doping levels of p- and n-Si(100) wafers were $2.7 \times 10^{15}$ - $1.5 \times 10^{16}$ $cm^{-3}$ and $1.6 \times 10^{18}$ - $8.0 \times 10^{18}$ $cm^{-3}$, respectively. The as-received Si(100) wafers were cleaned by standard RCA-1 treatment combined with HF dip, followed by a final step of using RCA-2 treatment to form a thin oxide layer on the surfaces. Next, the Si(100) substrates were thermally deoxidized in MOVPE reactor at 1000 °C and 950 mbar for 30 min.

The TBAs precursor source was used for surface termination by As. For the preparation of the A-type Si(100)-As 0.1° surface, TBAs flux was turned on at 450 °C – 670 °C for 15 min when the Si surface was oxidized, to achieve an arsenic ambient in the reactor. After turning off the TBAs, the Si(100) 0.1° surface was subsequently heated up to the deoxidation temperature at 1000 °C with arsenic back pressure. A quick cooling of the Si surface after 10 min to 420 °C under low $H_2$ pressure (~50 mbar) results in an As-modified, double-layer stepped A-type Si(100) 0.1° surface.

High offcut, Si(100) 4° surfaces were exposed to TBAs precursor at 830 °C. Cycles of TBAs exposure with a subsequent annealing in hydrogen and As background pressure (TBAs was turned off, pressure 950 mbar) between each cycle were applied to prepare As-terminated, double-layer stepped A-type Si(100) 4° surfaces [12].

The process route of B-type Si(100) 4° preparation was almost identical to that of the A-type Si(100) 0.1° surface but the reactor pressure was unchanged and kept at 950 mbar $H_2$ pressure during the cooling step from 1000 °C to 420 °C. A list of samples prepared for this work is described in Supporting information (SI).

GaP(100) films were epitaxially grown on the Si(100):As substrates. Triethylgallium (TEGa) and tertiarybutylphosphine (TBP) were used as precursors for the GaP growth. A two-step process consisting of a low temperature (~420 °C) nucleation and a relatively high temperature (~595 °C) growth of GaP layer was employed. During the nucleation and growth stages, the reactor pressure was 100 mbar and 50 mbar, respectively. GaP was nucleated on the Si(100):As surface by applying alternating pulses of TBP and TEGa (1 second each) at 420 °C. The grown GaP surface was cooled down to 420 °C under continuous TBP supply and annealed for 10 min without TBP supply in order to desorb excess phosphorus from the surface and to form a dimerized, P-rich reconstruction [20] of the GaP surface. The entire process was monitored by in situ RAS.

**2.2 XPS measurements**

After growth, samples were transferred in air to an XPS chamber and to the synchrotron facilities for HAXPES experiments. XPS measurements were performed on a Kratos Axis Supra spectrometer using monochromatic Al K$_\alpha$ radiation with an energy of 1486.58 eV (~ 1.5 keV hereafter). Survey and high-resolution spectra of the P 2p, Ga 3d, Si 2p and As 3d core levels were measured with an energy resolution of 0.45 eV (verified at the core level Ag 3d$_{5/2}$). High-resolution spectra were collected with a pass energy of 10 eV. The X-ray incidence angle and photoelectron emission angle were 54° and 0°, respectively, relative to the surface normal. The X-ray spot area was 0.21 mm$^2$. The atomic composition was calculated from the peak area intensities after subtracting the Shirley background using the ESCApe program (Kratos). Deconvolution of the core level spectra was performed using KolXPD software. The core-level peaks were fitted with Voigt profiles with fixed-width Lorentzian and variable-width Gaussian contributions. The full width at half maxima (FWHM) of bulk and interface were kept identical. Separation of doublet components and their ratio due to spin-orbit splitting was measured on bulk references [33] and used in the fit of heterostructure spectra.

**2.3 Gas cluster ion beam sputtering**

Commonly, Ar$^+$ monoatomic ion beam sputtering is used for the rapid erosion of metals and inorganic materials. Sputtering is associated with the destruction of atomic bonds and leads to the modification of the chemical states of atoms. Sputtering by Ar gas cluster ion beam (GCIB) is less destructive than by the monoatomic beam. During GCIB sputtering, clusters of atoms are formed by the adiabatic expansion of the working gas from high pressure through a nozzle into the gun (vacuum region). Atoms are then ionized by the impact of electrons and accelerated along the ion column. The column's interior separates the cluster ions according to their energy using a magnetic analyzer. The energy per ion can be tuned to a few electron volts only. The dissociated cluster on the surface removes material laterally from the near surface area and the subsurface area is sparsely disturbed [34]. GCIB sputtering in combination with *in situ* XPS was used to analyze the depth profile of organic materials [35], inorganic materials [36], and, in particular, GaP/Si(100) heterostructures [37].

GCIB sputtering was performed in an XPS chamber with an ion beam energy of 5 keV and a cluster size of 1000 ions. The ion beam current was 4.2 nA. The diameter of the cluster ion beams was 0.3 mm and the sputtered area was 1.5 x 1.5 mm$^2$. GCIB sputtering and *in situ* XPS measurements were performed continuously in cycles of 1, 2, 5 and 10 min intervals. GCIB-XPS studies were performed on the following samples: GaP films on n-Si(100):As 0.1º with different layer thickness of 16, 20, 25, and 50 nm, respectively, a GaP(100) wafer, and an oxidized n-Si(100):As substrate.

## 2.4 HAXPES measurements

HAXPES measurements of the GaP(As)/n-Si(100) 0.1° samples were carried out at the BL15XU beamline of SPring-8 [38,39] with an excitation energy of 5953 eV (~6 keV hereafter), using a hemispherical electron analyzer (VG Scienta R4000) with a pass energy of 200 eV, resulting in a total energy resolution of 240 meV at room temperature. The incident photons impinged with linear polarization collinear with the direction of emitted photoelectrons – the angle between incident photons and the electron analyzer entrance was fixed at 90°. An incidence angle of the X-ray beam was set to 2° from the surface plane.

A few samples with a GaP(As) layer thickness of 6, 8, 50 nm grown on p- and n-doped Si(100) substrates with 0.1º, 4º miscut orientations and prepared on A-type, B-type Si:As substrates were measured at the P22 beamline at DESY [40] with an excitation energy of 6 keV and a pass energy between 10 – 50 eV. Core level peaks were fitted similarly to that one used for XPS peak analysis. Results of DESY data analysis were included in SI.

## 3. Results and discussion

Fig. 1 a) shows the reflection-anisotropy (RA) spectrum measured on the surface of the Si (100):As-(1x2) A-type substrate. A peak at 3.2 eV indicates the presence of A-type (1x2) majority domains on the surface. In Fig. 1 b), RA spectra measured on the surfaces of 50 nm thick GaP films are shown: red line for two-domain GaP/n-Si(100):As 0.1° and green line for single-domain GaP(As)/n-Si(100):As 0.1°. The change of the RAS intensity due to the change of the domain ratio on the surface is indicated by an arrow.

In Fig. 1 c), the minority domain coverage was quantified from the RAS using a semiempirical model [41] and converges to ~10% for single-domain As-rich samples. Hereafter, we denote all As-rich samples with almost single-domain coverage as single-domain samples.

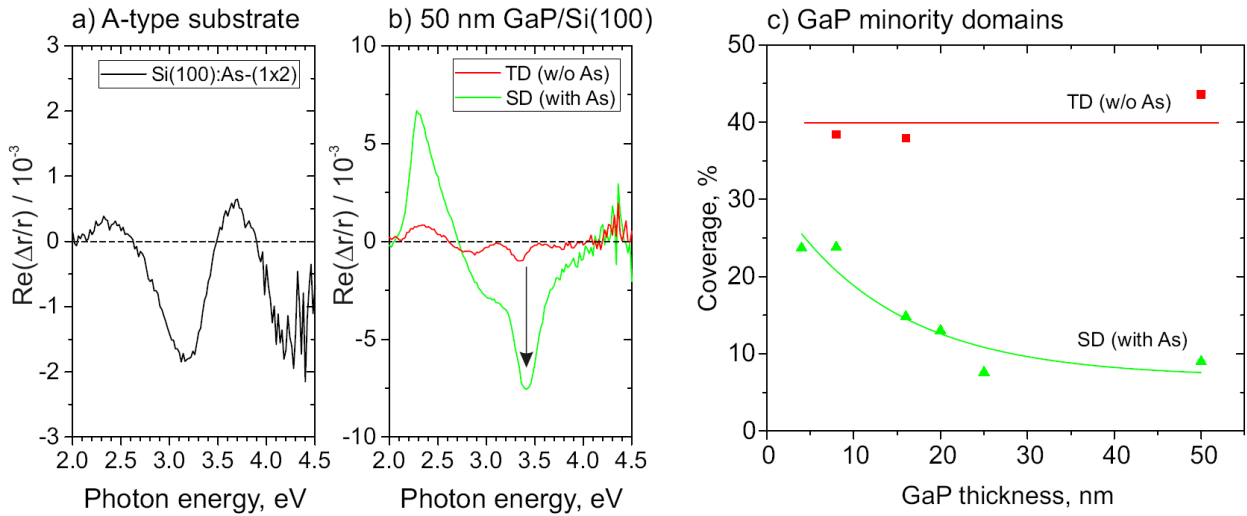

*Figure 1 – RA spectra of a) Si(100):As substrate reference with A-type domains on the surface, b) RA spectra of 50 nm thick single-domain GaP(As)/Si(100) (green) and two-domain GaP/Si(100) [29] (red) samples; equal numbers of domains cancel their local contribution to the RAS intensity for the two-domain sample. c) Coverage of the minority domains obtained from RA spectra.*

Figure 2 shows high-resolution XPS spectra of 4 nm thick single-domain GaP(As)/n-Si(100) 0.1° heterostructures measured at 90° and 20° emission angles, and relative differences in binding energies between components. Photoelectron spectra of bulk Si reference samples are included in SI [Fig. SI-1 a), b) and Tab. SI-T1]. The lower emission angle gives a higher surface sensitivity to the GaP layer and a higher interface sensitivity to the Si substrate. The experimental and deconvolved spectra are represented by symbols and solid lines, respectively. Two components of the P 2p core level have been resolved [Fig. 2 a) and b)]. $P_1$ is the bulk component, which is related to P-Ga bonds of the GaP overlayer. The second component $P_2$ is shifted by 0.6 eV (to higher binding energy) with respect to $P_1$. This component was already resolved in spectra measured on As-free 8 nm thick single-domain GaP/Si(100) samples by XPS and HAXPES and was identified as a contribution of Si-P bonds at the heterointerface: The $P_2$ area was 4% [29,31] and decreased with increasing surface sensitivity for As-free, 8-nm-thick sample (as expected for interfacial bonds). In Fig. 2 a), the area of $P_2$ is less than 2% for a 4-nm-thick As-rich sample and displays no dependence of the $P_2$ contribution on the emission angle. Therefore, the number of P-Si bonds is smaller for As-containing interfaces and $P_2$ seems to involve contributions of bonds in the GaP overlayer (possibly P-P bonds at antiphase boundaries). However, the area of $P_2$ is too small for a deeper analysis, as it is comparable to only a few percent of the error bars of the peak deconvolution analysis.

The Ga 3d peaks of Figs. 2 g) and h) contain $Ga_1$ bulk components (Ga-P bonds of the overlayer), an oxide component (Ga-O), and a component $Ga_0$ shifted by -0.3 eV (to lower binding energy). This component was previously assigned to Ga-Ga bonds [29,37] or Ga-Si bonds at Ga-polar interfaces [23]. In Fig. 2, the areas of $Ga_0$ and Ga-O components increase with decreasing electron emission angles. Origins of this component are therefore related to Ga-Ga bonds of Ga residuals [42] on the GaP surface and/or Ga-As bonds in the GaP overlayer.

In the Si 2p peaks [Fig. 2 c), d)], the $Si_1$ components originate from Si-Si bonds of the bulk Si(100) substrate. Another component, $Si_2$, is shifted by 0.3 eV with respect to $Si_1$. Their ratio $Si_2$ / $Si_1$ increases with decreasing emission angles. A similar chemical shift of 0.35 eV ± 0.2 eV and the dependence of the area ratio $Si_2$ / $Si_1$ on the emission angle were observed on the Si(100):As B-type (2x1) surface [12], where the component $Si_2$ was related to As-Si bonds on the Si surface. In Fig. 2, $Si_2$ evidently also originates from Si-As bonds at the buried GaP(As)/Si interface.

A chemical shift of 0.3 eV in Si 2p was also measured for P-polar, As-free GaP/Si heterostructures with P-Si bonds at the interface [29]. Therefore, the contribution of Si-P bonds to the $Si_2$ component originating from minority domains with a P-polar interface cannot be ruled out.

The As 3d peak of Figs. 2 e) and f) contains two components: $As_1$ with higher intensity and a lower-intensity component $As_2$ shifted by 0.5 eV. The As 3d peak partially overlaps with other peak contributions (see grey line at low binding energies) and was excluded from the deconvolution analysis. The area of $As_2$ decreases significantly with decreasing emission angles indicating that $As_1$ is a component related to the GaP(As) overlayer, while $As_2$ should originate from bonds at the buried interface.

The energy difference between As $3d_{5/2}$ (at ~40.9 eV) and Ga $3d_{5/2}$ (at ~18.9 eV) peaks is 22.0 eV for a GaAs(100) bulk crystal [43,44]. The energy difference between $As_1$ - $Ga_1$ is 21.8 eV measured in our GaP(As)/Si heterostructures. Therefore, $As_1$ is presumably associated to Ga-As bonds of the GaP(As) overlayer. The component $As_1$ was also measured on thicker GaP(As) overlayers without any contribution of interfacial bonds [compare Fig. 4 g)].

The As $3d_{5/2}$ peak binding energy of As-As bonds in metallic As is around 41.6 eV [43,44], i.e., the binding energy is shifted by 0.7 eV with respect to As $3d_{5/2}$ peak binding energy in GaAs (40.9 eV). The measured position of $As_2$ in Fig. 2 e) is shifted with respect to the Ga-As component $As_1$ by 0.5 eV. The shift of component $As_2$ thus coincides with the chemical shift of As-As bonds [localized at the interface according to Figs. 2 e), f)]. In addition to As-As bonds, the presence of Si-As bonds is also expected, since the surface of the Si(100) substrate was terminated by an As-(1x2) reconstruction prior to the GaP layers nucleation. At the moment, we cannot resolve these two contributions to $As_2$.

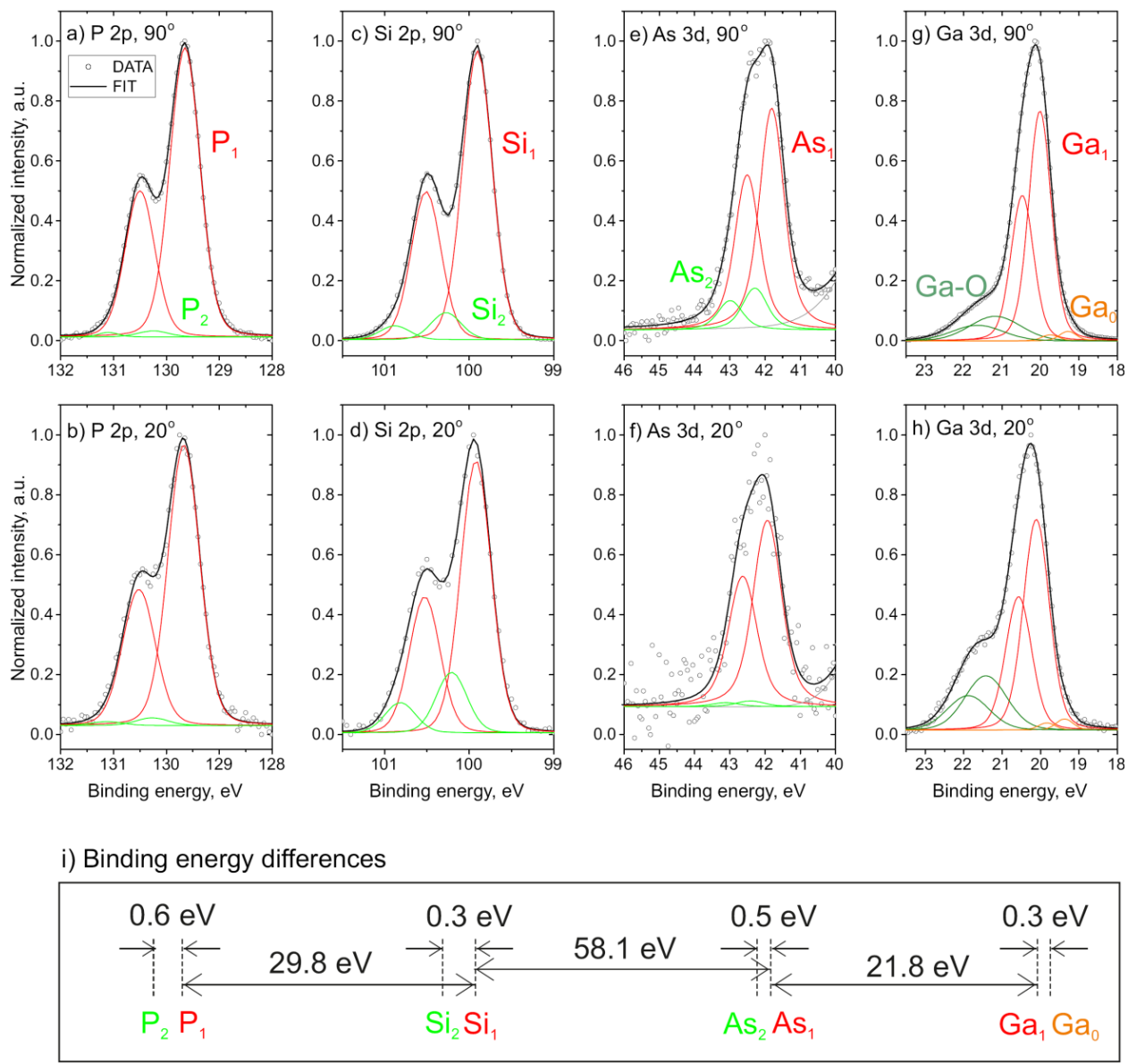

*Figure 2 – a)-h) XPS core level spectra of a 4 nm thick GaP overlayer prepared on As-terminated n-Si(100) and measured in normal emission geometry (90º) as well as in tilted geometry (20º). Experimental spectra and deconvoluted components are shown by symbols and solid lines, respectively. Identification of components can be found in the text. i) The relative differences in binding energy between the components (separations are not at the same scale).*

Fig. 3 displays depth profiles obtained by GCIB-XPS of the following samples: an As-terminated Si(100):As reference sample, covered with b) 16, c) 25, and d) 50-nm-thick GaP overlayers. Selected GCIB-XPS spectra for 16-nm-thick sample are shown in Fig. 4 (extended set of spectra is given in SI, Fig. SI-2, Tab SI-T3). The atomic concentration of the GaP overlayer was obtained by summing up the Ga and P atomic concentration profiles (see Fig. SI-5 in SI). Layers of contaminants on the surfaces consisting of C and O were significantly reduced by the sputtering procedure and their corresponding profiles were excluded from Fig. 3.

Fig. 3 a) shows that the arsenic concentration of the reference Si(100):As surface gradually decreased from 1.3 at.% to less than 0.1 at.% after 25 minutes of sputtering. Due to the preparation procedure, the arsenic is located on the surface. A true arsenic concentration profile would be represented as an abrupt step function. However, the measured profile is convoluted with the depth resolution function (DRF) of the spectrometer, which smears out the step function [45].

The GaP and Si profiles in Figs. 3 b), c) d) correlate: the GaP concentration decreases gradually with increasing sputtering time. The As concentration shows maxima located at the interface (~0.4 - 0.8 at.%) and a plateau of the As profile revealed As atom residues (0.2 - 0.3 at.%) in the GaP overlayer. Note, smaller concentration of As at interface in d) is related to decrease of depth resolution with overlayer thickness increase, i.e., with crater depth increase. Dependence of DRF on overlayer thickness is demonstrated in SI [Fig. SI-5 d)]. Estimated sputtering rates was determined by DRF maxima between 1.78 – 0.93 nm/min depending on the sample thickness.

Arsenic from the Si(100):As surface is, therefore, partially incorporated into the GaP lattice. Another source of As during the nucleation of GaP could be As precursor residuals in the MOVPE reactor. The As concentration has dropped below the "residual" concentration level in GaP at the end of the sputtering process.

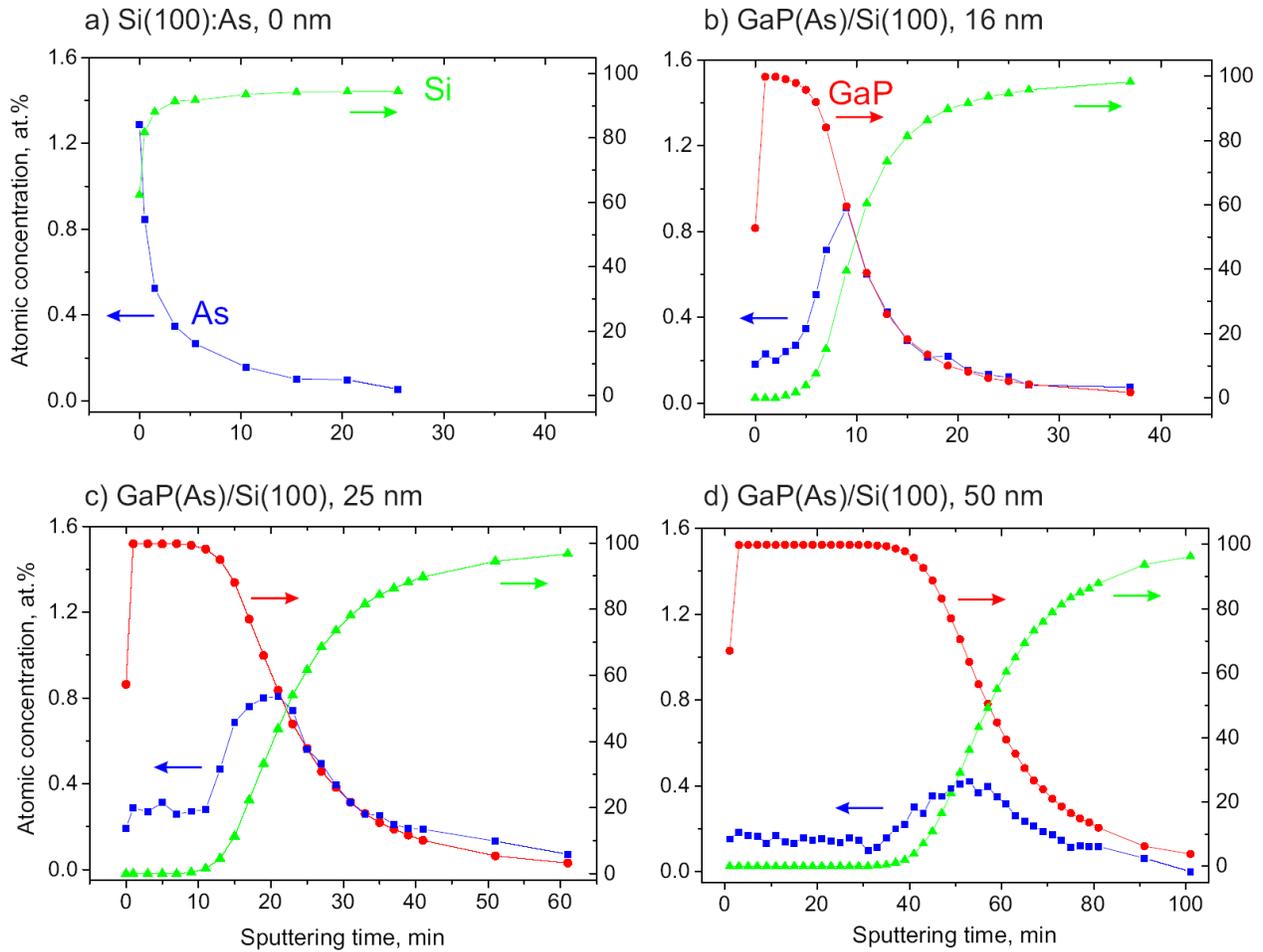

*Figure 3 – GCIB-XPS depth profiles of a) the reference Si(100):As sample, and b) 16, c) 25, d) 50-nm-thick GaP overlayers grown on top. As, Si, and GaP profiles are marked by blue, green, and red colors. Profile of GaP was obtained as a sum of P and Ga profiles. Arsenic residuals are present in the GaP overlayer, but mostly localized at the interface.*

Figure 4 shows a) Si 2p and c) As 3d GCIB-XPS core level spectra of a 16-nm-thick GaP(As)/Si(100) sample measured in normal emission geometry (90°) after 0 min – 37 min sputtering time. Selected deconvoluted b) Si 2p and d) As 3d spectra at 0 min (as-grown, without interface contribution), 5 min (with interface contribution), and 7 min (destructed interface) are also shown.

At the beginning of sputtering (0 min), there was no Si 2p peak, because the GaP overlayer thickness (16 nm) was larger than the information depth of XPS (the inelastic mean free path of Si 2p photoelectrons is λ = ~2.9 nm per TPP-2P formula [46]). The Si 2p peak began to appear at a sputtering time >3min. The position of the main Si 2p peak is preserved up to 7 min of sputtering [see dashed guide line in Fig. 4 a)], then the maxima shift to ~99.4 eV for larger sputtering times. This BE position should correspond to the disordered phase of silicon, $Si_{dis}$.

In Fig. 4 b), deconvoluted Si 2p spectra measured after 0 min, 5 min and 7 min of sputtering are shown. There are components $Si_1$ (bulk) and $Si_2$ (interface) with a chemical shift of 0.3 eV. The binding energies of $Si_1$ (99.9 eV) and $Si_2$ (~100.2 eV) are similar for the sputtered 16 nm and non-sputtered 4 nm samples [Fig. 4 b) and Fig. 2 c)]. A sum of $Si_1$ and $Si_2$ components therefore represents intensity contribution from the non-destructed buried substrate including interface, $Si_{sub}$ [filled yellow in Fig. 4 b].

In Fig. 4 b) a small $Si_{dis}$ component is resolved in 5 min Si 2p spectra (filled blue). Thus, despite a low sputtering rate, small fraction of disorder was induced by sputtering. Area of $Si_{dis}$ increased from 5 % of total intensity for 5 min Si 2p spectrum to 29 % for 7 min spectrum. The $Si_{dis}$ components should originate from the destructed Si crystal (amorphous Si). The $Si_{sub}$ and $Si_{dis}$ components were also resolved in Si 2p of 50-nm-thick sample after 39 min of sputtering with identical conditions (see Fig. SI-4 in SI).

The As 3d signal was observed on the as-grown sample, 0 min [Fig. 4 c), d)] and can be modelled with a single spin-orbit As 3d$_{5/2}$ and As 3d$_{7/2}$ doublet As$_1$. No interface [$As_2$, see Fig. 2 e)] or As oxide (with a binding energy between 44 – 46 eV) components were observed. The binding energy of As 3d$_{5/2}$ (41.8 eV) is the same as obtained for the non-sputtered 4-nm-thick sample in Fig. 2. It can thus be assumed that the origin of component $As_1$ is related to Ga-As bonds in the GaP(As) overlayer. The interface component $As_2$ [similar to one in Fig. 1e)] was not resolved, presumably because of the relatively low intensity of the As 3d peak. The As 3d peaks shift to lower binding energy immediately upon sputtering. In Fig. 4 d), As 3d 5 min spectrum contains two doublet components, $As_1$ and $As_{dis}$, which we attribute to contributions from the non-destructed GaP(As) layer and As in disordered layers. The intensity of component $As_{dis}$ in 7 min As 3d spectrum increases similarly to the increase of component $Si_{dis}$ in Si 2p. Released As by sputtering should react with disordered Si at interface.

GCIB-XPS data from overlayer was also measured. The P 2p and Ga 3d peaks shift to lower binding energy upon sputtering. Core level peak analysis results can be found in SI (Fig. SI-3 for 16 nm sample and Fig. SI-4 for 50 nm sample). Disordered shifted components dominate in overlayer core levels.

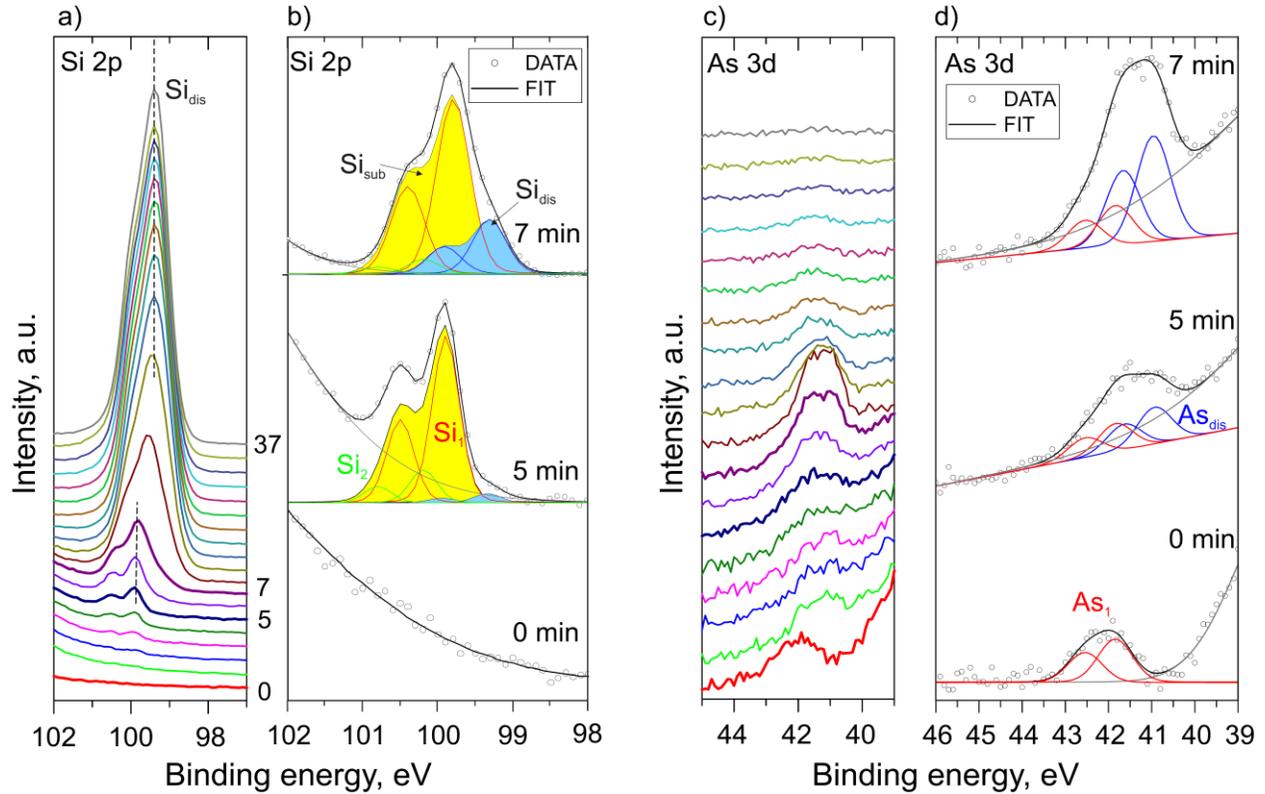

*Figure 4 – GCIB-XPS a) Si 2p, c) As 3d core level spectra of a 16-nm-thick GaP(As)/Si(100) sample. Sputtering time between 0 min and 37 min are indicated. The deconvoluted spectra of b) Si 2p and c) As 3d after 0 min. 5 min and 7 min of sputtering. Components are explained in the text.*

Synchrotron-based HAXPES offers non-destructive analysis of buried interfaces with an extended information depth compared to XPS. HAXPES has recently been applied to study VBOs and interface-induced built-in potentials in GaP/Si(100) heterostructures. In particular, strong band bending was measured in single-domain GaP/Si(100) heterostructures (P-rich interfaces) [29,33].

Figure 5 shows (a) P 2p, (b) Ga 3d, (c) Si 2p, (d) As $3p_{3/2}$ HAXPES spectra measured at SPring-8 on GaP(As)/Si(100) samples with an overlayer thickness 4 – 50 nm. Fits of P 2p, Ga 3d, and Si 2p core level peaks are in agreement with the corresponding XPS peak fits displayed in Fig. 3. In particular, the interface components $P_2$ and $Si_2$ with chemical shifts of 0.6 eV and 0.3 eV have been resolved. Note, HAXPES is proportionally less sensitive to the overlayer surface and, therefore, the oxide component is less pronounced in the Ga 3d spectra, whereas the $Ga_0$ component is resolved similarly to single-domain, As-free GaP/Si(100) heterostructures [29].

The intensity of the As 3d core level peaks was too low for detailed analysis. The intensity of the As $3p_{3/2}$ peak was weak but was distinguishable. In Fig. 5 c), As $3p_{3/2}$ peaks identified on GaP(As)/Si(100) heterostructures with an overlayer thickness between 4 nm and 50 nm are given. As $3p_{3/2}$ peak drops significantly with increasing the GaP(As) film thickness successively. As the spectra became noisier for increasing thickness of the GaP overlayers, only the 4 nm sample was selected for a peak deconvolution analysis. Two components were resolved in the As $3p_{3/2}$ peak: The component $As_2$ is shifted by 0.8 eV with respect to $As_1$. A similar shift of 0.5 eV was determined by XPS (Fig. 2). The difference of 0.3 eV is most likely related to a larger error bar caused by a low spectral intensity. Qualitatively, however, the deconvolution of As core levels agrees well for HAXPES and XPS. This is also true for the GaP bulk and Si substrate lines. There is only shift of the Si substrate and GaP emission lines for p-doped samples which are found to be shifted by about 0.2 - 0.4 eV for different experiments (see Tab. SI-T4 vs. Tab. SI-T5 in SI).

There are almost no shifts of core levels with sample thickness change in Fig. 5. In contrast to single-domain GaP/Si(100) samples without As, which show strong band bending and significant peak shifts [see Fig. 2c) in Ref. [29]], band bending appears to be eliminated in As-rich single-domain heterostructure on n-Si(100) substrate.

An estimate of the Fermi level position vs the valence band maximum lead to the following values: for n-Si(100), 0.1° wafer E($E_{VBM}$-$E_F$) = 1.1 ± 0.1 eV (see Fig. SI-1 d) in SI) and for GaP(100) overlayer E($E_{VBM}$- $E_F$) = 1.7 ± 0.1 eV (see Fig. 7 below). The deduced values of the Fermi level position which is close to the conduction band minima (CBM) for Si and showing weak n-doping (n⁻) for the grown GaP layer also follows the measured core level binding energies and the known binding energy differences between core level positions and VBM.

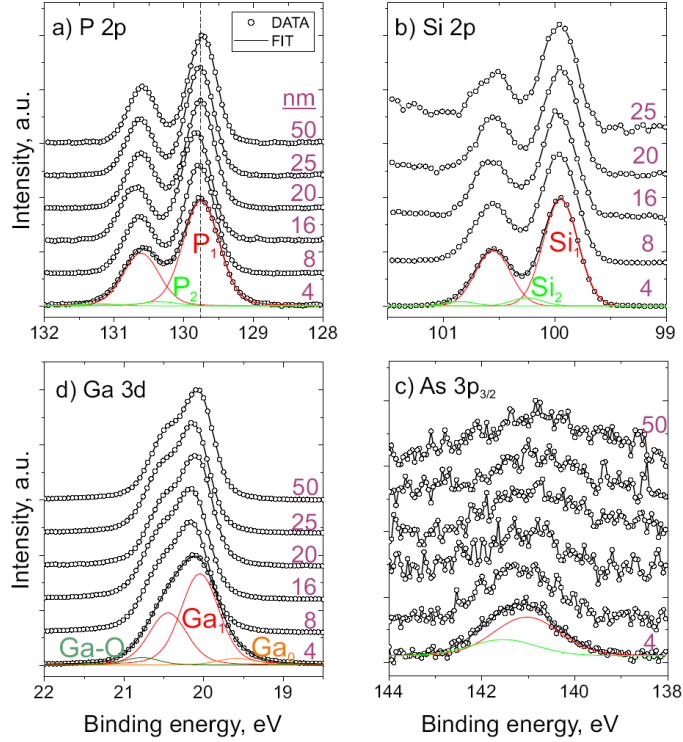

*Figure 5 – HAXPES spectra measured on GaP(As)/Si(100) samples with a GaP film thickness of 4 nm – 50 nm: a) P 2p, b) Si 2p, c) As 3d, d) Ga 3d spectra. The notations of the individual components are similar as given in Fig. 2. The interface components (green) were observed in the P 2p, Si 2p, As $3p_{3/2}$ peaks.*

Based on the observation of components $As_1$, $As_2$ and $Si_2$, we propose a heterostructure model consisting of P-dimers on surface, $Ga_{1-x}As_xP$ (overlayer, x<0.01) and Ga-As-Si (interface) on top of Si(100) substrate. Such a sequence of layers is supported by RAS measurements: there was no change of sign of the RAS signal between the A-type Si(100):As-(1x2) and A-type GaP(100)-(1x2) spectra. The component $P_2$ can be explained by the presence of antiphase domain boundaries or fractions of domains with P-polar interface. This leads to the conclusion that the surface of the Si(100) substrate prior to the GaP nucleation may contain a mixture of As-As, As-Si, or even Si-Si dimers. Correspondingly, As atoms on the Si(100):As-(2x1) surface may be released during the nucleation procedure and diffuse into the GaP layer. As a result, Ga-As-Si bonds at the interface are not predetermined. In such heterostructures, the interface charge and band bending may be compensated by the coexistence of cation-Si and anion-Si atomic pairs.

The VBO between the GaP(As) bands and the Si substrate bands was derived by measuring the core level binding energies and the VBM in heterostructures and the corresponding bulk references [47,48]:

$$\Delta E_V^{exp} = \Delta E_{Si}^{bulk} - \Delta E_{GaP}^{bulk} + \left(E_{GaP}^{Het} - E_{Si}^{Het}\right) \quad (1)$$

where bulk constants $\Delta E_{Si}^{bulk} = E_{Si2p}^{bulk} - E_{VBM}^{bulk} = 98.82$ eV is the mean energy difference between Si 2p and the VBM of bulk Si(100) wafers and $\Delta E_{GaP}^{bulk} = E_{GaP}^{50\,nm} - E_{VBM}^{50\,nm}$ is the energy difference of the bulk GaP(100) films (128.03 eV for P $2p_{3/2}$ and 18.36 eV for Ga $3d_{5/2}$ core levels). Mean values have been used for the final evaluation of the energy diagrams which should be considered as a mean value with the deviation as given from the experiments. A list of CL – VBM energy difference is given in SI (Tab SI-T2). The last term in Eq. 1 is the difference in core level binding energy measured from heterostructure (Het) samples.

The applicability of photoelectron spectroscopy for VBO measurements is governed by the escape depth of photoelectrons from the buried interface and substrate. Heterostructures with an overlayer thickness below 10 nm can be measured by XPS. The information depth can be extended up to ~30 nm by HAXPES when utilizing ~6 keV photons [29].

Figure 6 shows the experimental VBO values on GaP(As)/n-Si(100) 0.1°, A-type samples with different overlayer thickness experimentally identified by XPS and HAXPES. VBO values between 0.58 - 0.67 eV ± 0.05 eV were obtained utilizing these two different experimental approaches. The extema of this VBO range are determined by the XPS-derived values, while the VBO values derived from HAXPES show a significantly reduced variability (0.61 - 0.64 eV). This can be explained by the higher surface sensitivity of the XPS measurements, making them more susceptible to surface and interface effects, such as band bending, film strain, presence of contamination, and/or different composition.

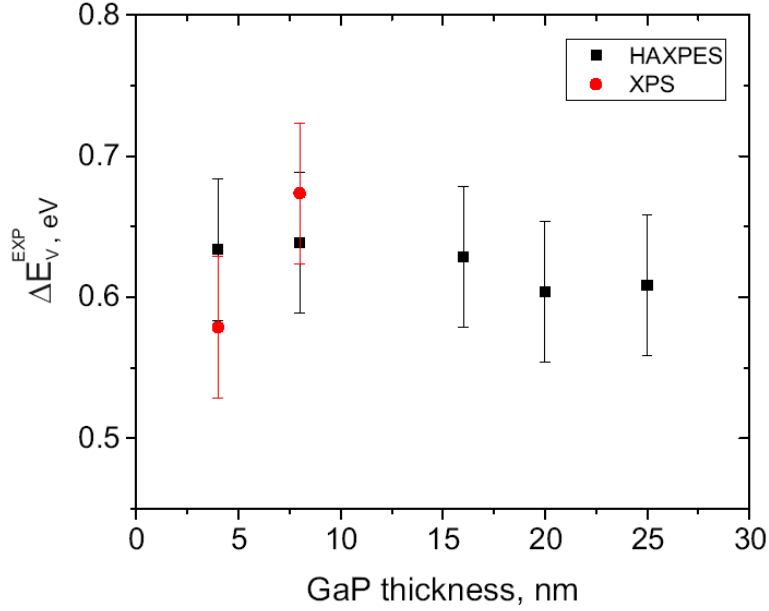

*Figure 6 – Comparison of measured VBOs of the GaP(As)/n-Si(100), 0.1º miscut, A-type samples by HAXPES and XPS. A good agreement of the VBO values obtained at the synchrotron and in the laboratory is observed.*

The VB HAXPES spectra and energy diagrams of the n⁻-GaP(As)/n-Si 0.1° , A-type samples are shown in Fig. 7. In a), VBM value of 1.7 eV was derived for the 50 nm sample. The valence band edge was shifted by ~0.7 eV from the edge of the Si(100):As (0 nm) reference sample. This shift is similar to the value of the measured VBO of ~0.6 eV using core level positions (Eq. 1). Moreover, the position of the Si VB edge in the heterostructure (~1.1 eV) is slightly higher than the VB edge position of the Si(100):As sample (1.0 eV) and the same as the VBM edge position of the reference n-Si(100) wafer (1.1 eV for HAXPES, see Tab. SI-T1 in SI), indicating that the Si at the interface remains strongly n-type.

In Fig. 7 b), flat-band diagram for n-Si substrate samples is shown. The measured values of VBOs and VBM are shown in red. Taking into account the corresponding bulk band gaps, the CBO, $\Delta E_C = 0.6$ eV was deduced (similar to the VBO value).

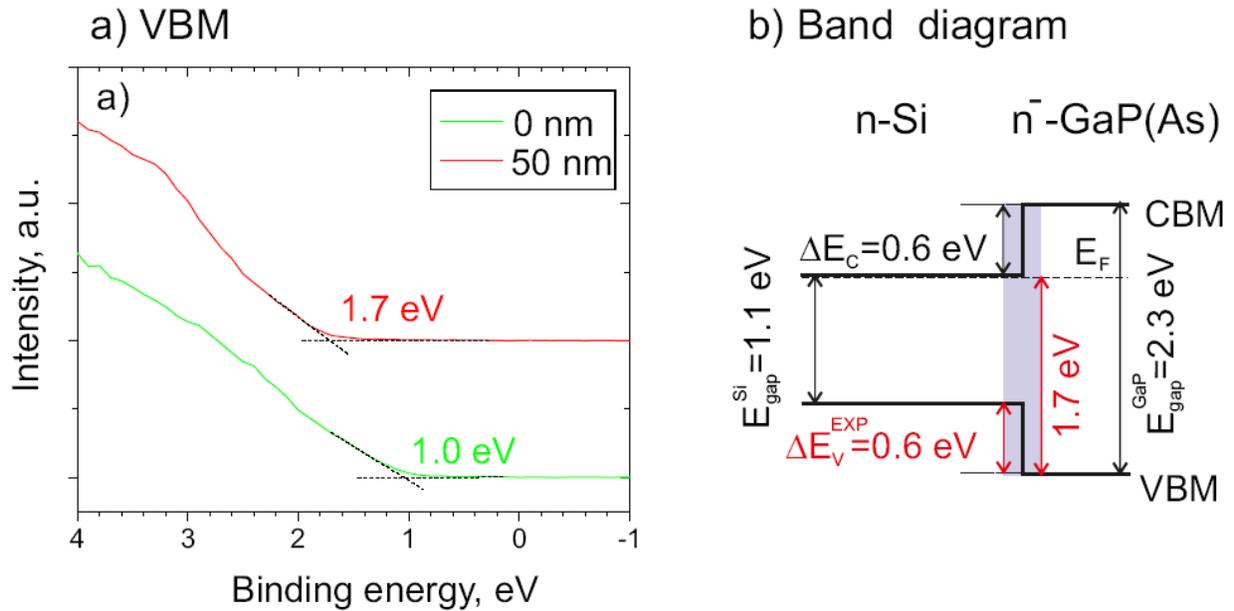

*Figure 7 – a) Valence band HAXPES spectra of the As:n-Si(100) (0 nm) and GaP(As)/n-Si(100) (50 nm) samples. b) Band alignment diagram. Measured values of VBO and VBM are marked in red. A near interface region, which is assessable by the photoemission measurements, is schematically indicated by violet area. The Fermi level at ~1.7 eV above the GaP VBM lies in a proximity of the conduction band minimum (CBM).*

For the p-doped Si(100) wafer, the Fermi level was found at 0.6 eV [see Fig. 1 d) and Tab. SI-T1 in SI]. For the p-doped 50-nm-thick GaP(As)/p-Si heterostructure, the Fermi level position of the GaP overlayer is also slightly shifted to lower values: the VBM of GaP is 1.3 eV on p-Si substrates [see Fig. SI-7 a) and Tab SI-5 in SI] and 1.7 eV on n-Si substrates. Therefore, an upward band banding in the GaP overlayer and a downward band bending in Si substrate toward interface were suggested for p-type samples [see Fig. 7 b) in SI]. For the same substrate type (i.e., A-type Si(100) 0.1° surface) the HAXPES-derived VBO of GaP(As)/p-Si(100) heterostructures were found to be $0.63 \pm 0.1$ eV and thus similar to that of the n-Si samples.

Possible As interdiffusion into p-Si substrate [49] may change band bending directions at the interface: Arsenic is an n-type dopant for Si and charge compensation between interface states and doping states should occur. In SI [Fig. SI-7 c)], we have therefore also suggested an alternative model of band alignment in case of As interdiffusion into the p-Si substrate. A more detailed analysis of such interfaces was not carried out in the present paper due to lack of experimental data of p-type Si samples.

Our data show that the VBO is just given by the difference of the $E_{VBM}$- $E_F$ values after aligning the Fermi levels of the Si substrate and the GaP overlayer across the interface. The measured VBO values of GaP(As)/Si(100) samples do not depend much (0.6 ± 0.1 eV) on the type of doping of the Si substrate, Si substrate miscut angle, and type of As-terminated surface (see Tab. SI-T5 in SI).

Evidently, there could be Fermi level pinning, which is related to Si substrate surface bonds or bonds at buried Si interface. Such bonds could form bandgap states leading to Fermi level pinning positions at the observed energy range. In the valence band spectra, we found indication of the emission from the bandgap states (see Fig. SI-9 in SI), but measured intensities were very weak and additional verification of emission from the bandgap states is needed. Therefore, we suggest that the localized interface electronic states may play a crucial role in charge displacements and band alignment at the heterojunction.

**Conclusions**

The atomic composition of GaP(As)/Si (100) heterostructures with a thickness of 4 nm – 50 nm prepared by MOVPE was investigated by photoelectron spectroscopy methods in combination with depth profiling using Ar GCIB. The preparation of a clean Si(100) surface in an As environment allows the growth of an almost single-domain GaP(100) films on Si(100) surfaces modified with As. The depth profile analysis revealed the localization of As at the buried GaP/Si:As(100) interface, and, in addition, dissolving of As atoms in the GaP lattice. Based on the analysis of core level spectra, the interface structure was proposed: The heterostructure consists of a mostly Ga-polar interface with an interfacial Ga-As-Si layer stacking sequence. A fraction of domains with P-polarity and a Ga-P-Si layer sequence at the interface is also present. Similar VBOs of 0.6 ± 0.1 eV were obtained on GaP(As)/Si(100) heterostructures with variable substrate doping type, substrate miscut orientations or As surface type on Si substrate. It is interesting to note that the alignment of the buried junction Si(100)/(As)/GaP interface could be investigated by a number of different experimental approaches leading to very consistent results.

**Acknowledgments**

We gratefully acknowledge financial support by the Czech Science Foundation (GACR, proj. no. 18-06970J), by the German Research Foundation (DFG, proj. no. HA3096/10-1 and PAK 918), and the Federal Ministry of Education and Research (BMBF project H2Demo). The work is also supported by Operational Program Research, Development and Education financed by European Structural and Investment Funds and the Czech Ministry of Education, Youth and Sports (Project No. SOLID21 - CZ.02.1.01/0.0/0.0/16_019/0000760) and NIMS microstructural platform as a program of "Nanotechnology Platform" (Project No. 12024046) of MEXT, Japan. The HAXPES measurements at SPring-8 were performed with the approval of NIMS Synchrotron X-ray Station (Proposal Nos. 2018A4908, 2018B4909, and 2019A4910). We acknowledge DESY (Hamburg, Germany), a member of the Helmholtz Association HGF, for the provision of experimental facilities. Beamtime at DESY was allocated for proposal I-20190961. Funding for the HAXPES instrument at beamline P22 by the Federal Ministry of Education and Research (BMBF) under framework program ErUM is gratefully acknowledged. The access to the MetaCentrum computing facilities provided under Project No. LM2010005 funded by the Ministry of Education, Youth, and Sports of the Czech Republic is highly appreciated.

# Supporting Information

**Combining advanced photoelectron spectroscopy approaches to analyse deeply buried GaP(As)/Si(100) interfaces: Interfacial chemical states and complete band energy diagrams**

O. Romanyuk[1,*], A. Paszuk[2], I. Gordeev[1], R.G. Wilks[3,4], S. Ueda[5,6,7], C. Hartmann[3], R. Félix[3], M. Bär[3,4,8,9], C. Schlueter[10], A. Gloskovskii[10], I. Bartoš[1], M. Nandy[2], J. Houdková, P. Jiříček[1], W. Jaegermann[11], J.P. Hofmann[11], T. Hannappel[2]

[1]*FZU – Institute of Physics of the Czech Academy of Sciences, Prague, Czech Republic*
[2]*Dep: Fundamentals of Energy Materials, Institute of Physics, Ilmenau University of Technology, Ilmenau, Germany*
[3]*Department Interface Design, Helmholtz-Zentrum Berlin für Materialien und Energie GmbH, Berlin, Germany*
[4]*Energy Materials In-Situ Laboratory Berlin (EMIL), Helmholtz-Zentrum Berlin für Materialien und Energie GmbH, Berlin, Germany*
[5]*Synchrotron X-ray Station at SPring-8, National Institute for Materials Science (NIMS), Hyogo, Japan*
[6]*Research Center for Functional Materials, NIMS, Tsukuba, Japan*
[7]*Research Center for Advanced Measurement and Characterization, NIMS, Tsukuba, Japan*
[8]*Department of Chemistry and Pharmacy, Friedrich-Alexander-Universität Erlangen-Nürnberg, Erlangen, Germany*
[9]*Helmholtz-Institute Erlangen-Nürnberg for Renewable Energy (HI ERN), Berlin, Germany*
[10]*Deutsches Elektronen-Synchrotron DESY, Ein Forschungszentrum der Helmholtz-Gemeinschaft, Hamburg, Germany*
[11]*Surface Science Laboratory, Department of Materials and Earth Sciences, Technical University of Darmstadt, Darmstadt, Germany*

*corresponding author, e-mail: romanyuk@fzu.cz

**Binding energies of bulk references**

X-ray photoelectron spectroscopy (XPS) spectra measured on oxidized commercial Si(100) wafers by Al Kα excitation energy (1486.58 eV) are shown in Fig. SI-1 a), b). Hard X-ray photoelectron spectroscopy (HAXPES) spectra measured with 6 keV are shown in Fig. SI-1 c), d). The corresponding binding energies are present in Tab. SI-T1. Valence band edges were extrapolated from the slope of the VB spectra. Note, the valence band maxima (VBM) measured by XPS on n-Si sample is higher than the band gap value of Si (1.1 eV). Therefore, XPS spectra were shifted by the surface band bending. The relative differences between the core level peak position and VBM are similar for XPS and HAXPES (see Tab. SI-T2).

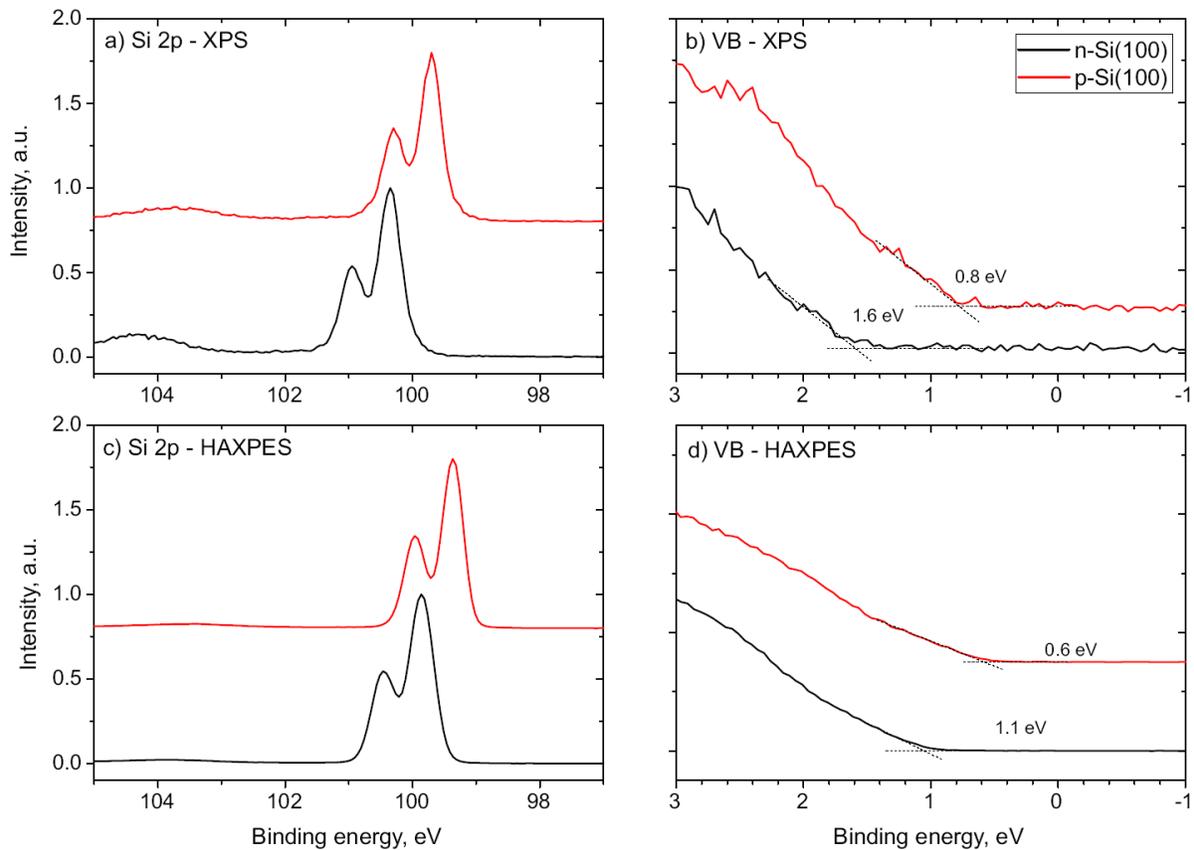

*Figure SI-1 – a), c) Si 2p core level and b), d) valence band photoemission spectra of bulk (oxidized) Si(100) wafers measured by XPS and HAXPES.*

*Table SI-T1 – Binding energies of bulk Si(100) wafer references measured by XPS and HAXPES.*

| Substrate type | BE(Si 2p$_{3/2}$), eV | $E_{VBM}$-$E_F$, eV | Source |
|---|---|---|---|
| n-Si(100) wafer, 0.1° | 100.35 | 1.6 | XPS |
| p-Si(100) wafer, 2° | 99.70 | 0.8 | XPS |
| n-Si(100) wafer, 0.1° | 99.86 | 1.1 | SP-8 |
| p-Si(100) wafer, 2° | 99.37 | 0.6 | DESY/SP-8 |

In Tab. SI-T2, Binding energy difference between core levels and VBMs are given for Si(100) and 50-nm-thick GaP(As) reference samples. Mean values of $\Delta E_{Si}^{bulk} = E_{Si2p}^{bulk} - E_{VBM}^{bulk} = 98.82$ eV, $\Delta E_{P2p}^{bulk} = E_{P2p}^{50\,nm} - E_{VBM}^{50\,nm} = 128.03$ eV and $\Delta E_{Ga3d}^{bulk} = E_{Ga3d}^{50\,nm} - E_{VBM}^{50\,nm} = 18.36$ eV were used for valence band offset (VBO) calculation by Eq. 1 in paper. It has been verified that valence band offset variation below 0.1 eV is possible without averaging of bulk constants.

*Table SI-T2 - Binding energy differences between core level and VBM for bulk Si(100) wafers and 50-nm-thick GaP(100) reference samples.*

| | Sample | CL - VBM difference, eV | | |
|---|---|---|---|---|
| | | BE(Si 2p - $E_{VBM}$) | BE(P 2p - $E_{VBM}$) | BE(Ga 3d - $E_{VBM}$) |
| HAXPES | n-Si wafer | 98.80 | - | - |
| | p-Si wafer | 98.80 | - | - |
| | p-Si:As_A-type, 0.1 | 98.77 | - | - |
| | n-Si:As_A-type, 0.1 | 98.85 | - | - |
| | p-Si:As_A-type, 4 | 98.84 | - | - |
| | p-Si:As_B-type, 4 | 98.80 | - | - |
| | 50nm GaP(As)/n-Si | - | 128.02 | 18.35 |
| | 50nm GaP(As)/p-Si | - | 128.05 | 18.37 |
| XPS | n-Si wafer | 98.76 | - | - |
| | p-Si wafer | 98.94 | - | - |
| | Average | 98.82 | 128.03 | 18.36 |
| | Err. Bar | 0.06 | 0.02 | 0.01 |

**GCIB-XPS intensities**

In Fig. SI-3, measured GCIB-XPS core level spectra are shown. Core level peak positions do not depend much on overlayer thickness. In case of Si substrate, positions of the Si 2p core levels measured at non-destructed buried interface are also similar for all samples. Flat band alignment was deduced by GCIB-XPS.

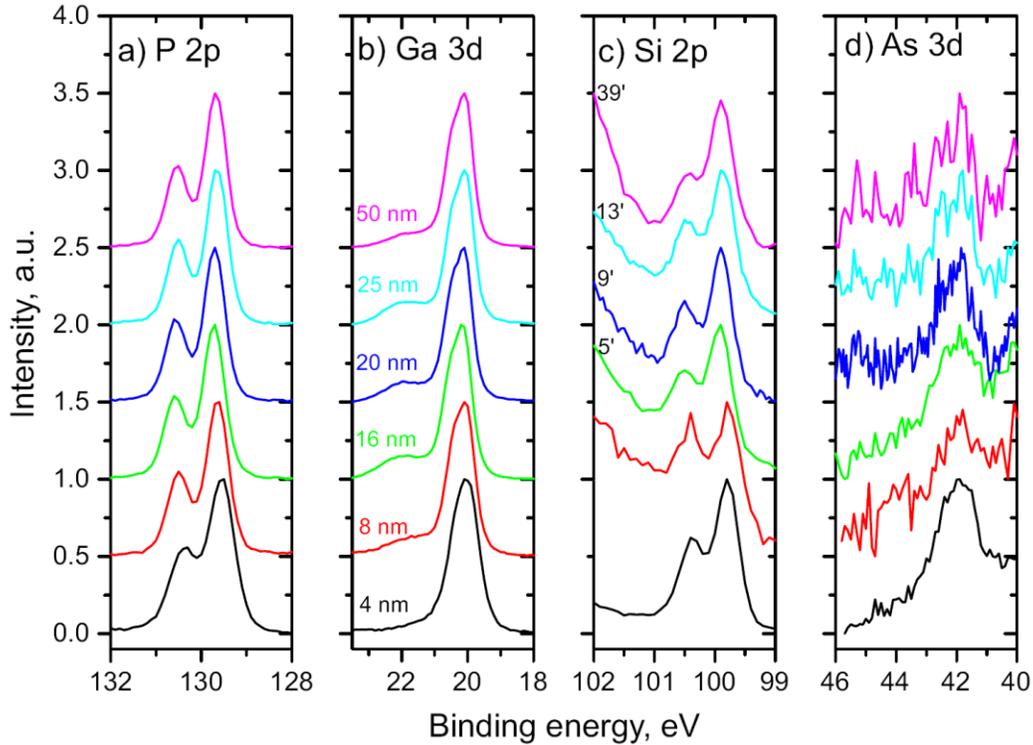

*Figure SI-2 – GCIB-XPS spectra of a) P 2p, b) Ga 3d, and d) As 3d core levels measured on as-grown, non-sputtered 4 – 50-nm-thick GaP(As)/n-Si(100) samples. In c), Si 2p spectra were measured on non-sputtered 4 nm, 8 nm samples and after 5, 9, 13, and 39 min of sputtering for 16, 20, 25, 50 nm samples, correspondingly.*

In Tab. SI-T3, GCIB-XPS core level binding energies and obtained VBOs values are given for GaP(As)/n-Si(100) samples. Data were obtained from as-grown, non-sputtered samples (black numbers), except of energies at non-destructed buried interfaces (blue numbers), which were obtained by sputtering.

*Table SI-T3 – GCIB-XPS binding energies measured on non-sputtered (black numbers) and sputtered (blue numbers) GaP(As)/n-Si(100) sample surfaces. The corresponding spectra are shown in Fig. SI-2. The VBO values derived by P 2p and Ga 3d core levels were averaged, <VBO>, and shown in Fig. 6 in the paper.*

| GaP thickness, nm | Substrate type | BE(P $2p_{3/2}$), eV | BE(Ga $3d_{5/2}$), eV | BE(Si $2p_{3/2}$), eV | BE(As $3d_{5/2}$), eV | VBO(P2p) | VBO(Ga3d) | <VBO>, eV | Source |
|---|---|---|---|---|---|---|---|---|---|
| 4 | n-type, 0.1°, A-type | 129.54 | 19.93 | 99.78 | 41.70 | 0.55 | 0.61 | 0.58 | XPS |
| 8 | | 129.64 | 20.00 | 99.77 | 41.80 | 0.66 | 0.69 | 0.67 | XPS |
| 16 | | 129.72 | 20.12 | 99.90 | 41.85 | - | - | - | XPS/GCIB |
| 20 | | 129.66 | 20.05 | 99.86 | 41.83 | - | - | - | XPS/GCIB |
| 25 | | 129.66 | 20.05 | 99.88 | 41.80 | - | - | - | XPS/GCIB |
| 50 | | 129.68 | 20.06 | 99.90 | 41.80 | - | - | - | XPS/GCIB |

Fig. SI-3 shows the dependence of the GCIB-XPS intensities of P 2p and Ga 3d core levels on the sputtering time measured on a 16 nm thick GaP(As)/n-Si(100) sample. The Si 2p and As 3d spectra are given in the paper. The sputtering time between 0 – 37 min is indicated. The GaP overlayer was continuously sputtered away. The line shapes of the P 2p and the Ga 3d peaks were changed by sputtering. In particular, peak line shapes are different for P 2p after 0 min and after sputtering, >1 min. A disordered phase has been formed on the GaP surface during sputtering. After the first sputtering step (1 min), the peak positions shifted towards the low binding energy side. The shift is caused by the removal of the contamination layer and formation of the disordered phase with different Ga/P stoichiometry and bonding environment.

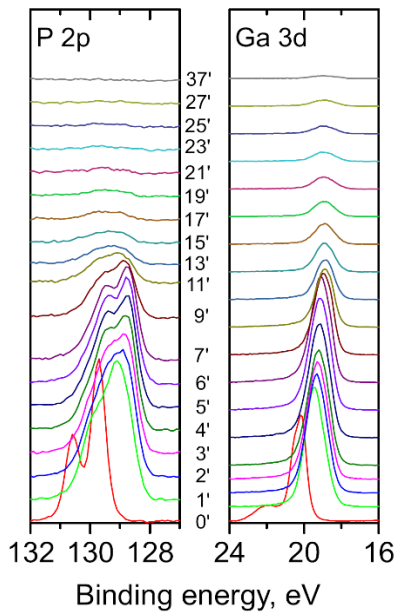

*Figure SI-3 – The dependence of GCIB-XPS core level intensities measured on 16 nm thick GaP(As)/Si(100) sample during sputtering time between 0 – 37 min.*

Figure SI-4 shows selected core level GCIB-XPS intensities measured on 50-nm-thick GaP(As)/Si(100) sample. A schematic model of the sputtering process is shown on the left hand side (deduced from the deconvoluted intensities). Sputtering times of 0, 5, and 39 min correspond to as-grown sample (without interface contribution), sputtered overlayer (without interface contribution), and sputtered heterostructure close to interface (with interface/substrate contribution), correspondingly. The 0 min spectra of the as-grown samples contained one bulk component $P_1$ in P 2p, whereas the interface component $P_2$ is missing; one $As_1$ component in the As 3d peak (As bonds in the GaP lattice); the bulk component $Ga_1$, oxide and $Ga_0$ components in Ga 3d. The thickness of the sample was larger than the XPS information depth and, therefore, no intensity from the interface and the Si substrate was present, i.e. no Si 2p intensity was observed. Schematic sputtering model is also given in Fig. SI-4.

The 5 min P 2p spectrum contains three doublet components: two of them $P_{dis}$ should be related to the disordered GaP(As) overlayer since there was still no Si 2p peak. One doublet component $P_1$ corresponds to P in bulk GaP (with the same BE position as $P_1$ in 0 min P 2p spectrum). The $P_1$ intensity is lower than $P_{dis}$, therefore, disorder layer has to be thicker and lie on a top of ordered phase in the bottom of crater (see scheme on a left hand side). A sum of $P_1$ and $P_{dis}$ gives a P 2p reference spectrum for the sputtered overlayer (filled by violet).

The 5 min As 3d peak contains two doublets: $As_1$ component and additional $As_{dis}$ component. Similar disordered component $As_{dis}$ was observed on sputtered 16-nm-thick sample (see Fig. 4 in the paper). The 5 min Ga 3d spectrum contains small $Ga_1$ bulk component, shifted disordered component $Ga_{dis}$, and an additional component $Ga_m$ of metallic Ga appeared (due to sputtering-induced effect) [1]. The Ga oxide was sputtered away and, therefore, Ga-O component is disappeared. Line shape of Ga 3d spectra after 5 min and 39 min of sputtering is similar except of increasing concentration of metallic phase $Ga_m$.

The 39 min P 2p spectra has a different line shape than the 5 min P 2p spectra. In order to deconvolute the spectrum, we used a reference line shape of the 5 min P 2p spectrum (consisting of disordered overlayer contribution and residual intensity from the ordered phase, $P_1+P_{dis}$) and introduced one more doublet component, $P_{int}$. This 'extra' intensity correlates to the interface or substrate: it appeared together with the Si 2p signal. Binding energy of $P_{int}$ is 129.5 eV, which is shifted in respect to the interface component $P_2$ in P 2p (130.2 eV, ordered phase, Fig. 2 a)].

Therefore, $P_{int}$ could be identified as a contribution from the disordered P-Si bonds close to the interface, i.e. sputtering-induced atomic intermixture at interface. By other words, we were not able to get the non-destructed overlayer interface components at buried interface by GCIB-XPS.

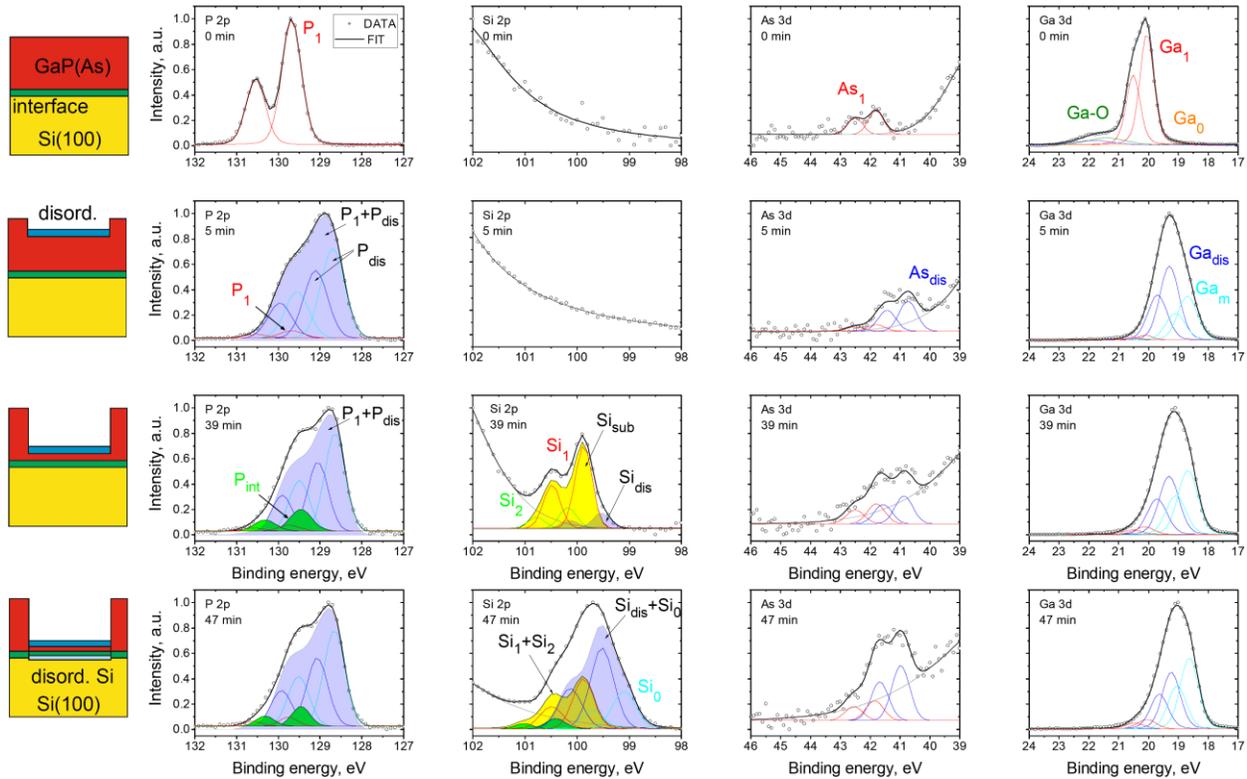

*Figure SI-4 – Si 2p spectra measured on a 50 nm thick GaP(As)/n-Si(100) heterostructure by GCIB-XPS. a) Dependence on the sputtering time from 33 min to 65 min in steps of 2 min. b) Deconvolution of the Si 2p peak measured from non-destructed interface after 39 min of overlayer sputtering.*

In contrast to disturbed overlayer core levels, the Si 2p spectrum from the substrate after 39 min of sputtering consists of non-destructed bulk component $Si_1$, interface component $Si_2$, and a small intensity component from the sputtering-induced disordered Si, $Si_{dis}$. The same components were observed in the deconvoluted Si 2p spectrum from 16-nm-thick sample.

The 39 min As 3d spectrum contains two components as 5 min spectrum. However, component ratio has changed close to interface: the contribution from the built-in As in GaP lattice, $As_1$ contribution has increased close to interface. It agrees with increased concentration of As at interface according to depth profiles in Fig. SI-6. Component ratio of ~50:50 for $As_1/As_{dis}$ was also measured after 5 min of sputtering of 16-nm-thick sample in Fig. 4 d).

**Concentration profiles**

Figure SI-5 shows the derived concentration profiles from a) 16 nm, b) 25 nm, and c) 50 nm thick GaP(As)/n-Si(100) heterostructures (symbols) and bulk GaP(100) reference (solid lines). The measured profiles of Ga (blue squares) and P (red circles) in heterostructures follow the corresponding profiles in the reference until the Si contribution appears. Then the concentrations of Ga and P decrease and the atomic concentration of Si increases.

Higher concentration of Ga than P during sputtering indicates preferential sputtering of anion atoms, i.e. P [2]. This effect has previously been observed on other compound semiconductors: Cation atoms accumulates on the surface during sputtering [1]. Therefore, Ga profile is not parallel to the P profile during sputtering of GaP. In the paper, we summed up Ga and P profiles for clarity (despite stoichiometry of GaP has changed) to obtain the remaining GaP overlayer profile [Fig. 3 b)-d)].

The slope of the GaP and Si profiles varies with the thickness of the overlayer [Fig. SI-6 a)-c)]. This can be explained by the dependence of the depth resolution function (DRF) on the overlayer thickness [3–5] (rather than by smearing of the Si interface). In Fig. SI-5 d), the DRF functions are shown. They were obtained by differentiation Si profiles (the differentiation of the integrated GaP profiles gave the same result). The dependence of the shape of the DRF on the overlayer thickness is obvious. Thus, the shape of the obtained DRFs is not a constant for GCIB sputtering, but strongly depends on the depth of the crater.

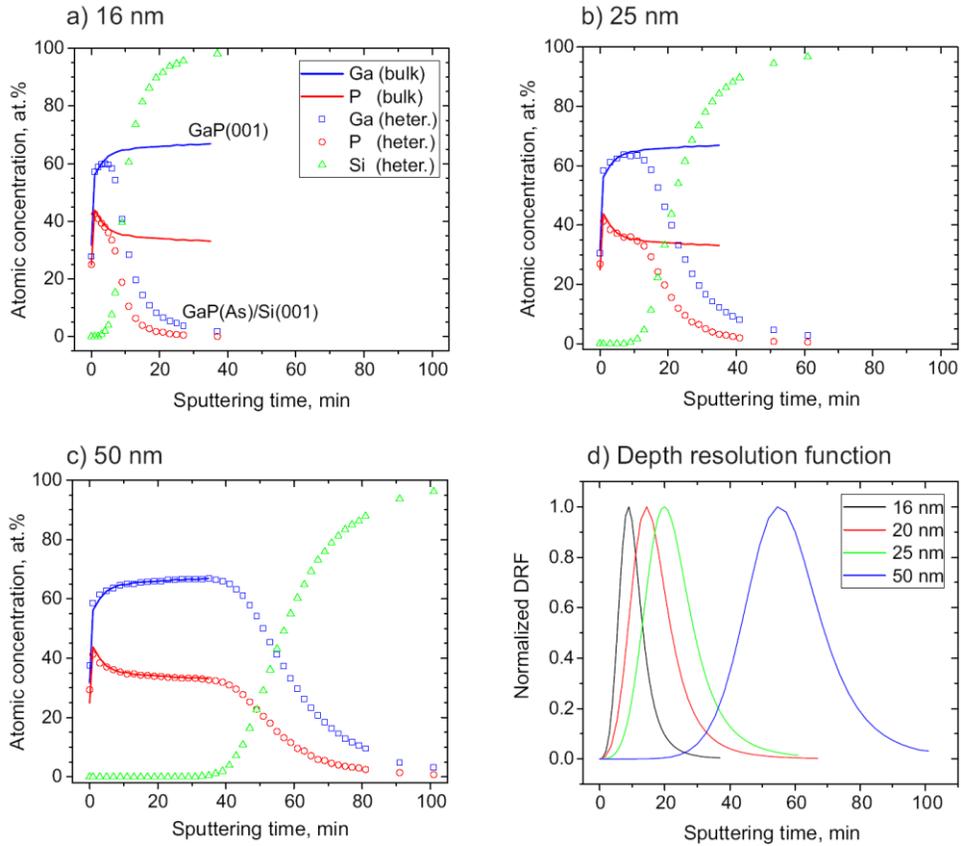

*Figure SI-5 – Depth profiles of atomic concentration measured on a) 16 nm, b) 25 nm, and c) 50 nm thick GaP(As)/Si(100) samples by in-situ GCIB-XPS. Concentration profiles of heterostructures and reference samples are indicated by symbols and solid lines, respectively. The lower concentration of P confirms the preferential sputtering of P. The As profiles are shown in the manuscript (see Fig. 3). d) Normalized depth resolution functions obtained for GCIB sputtering.*

The peak position maxima of the DRF lie at $t_{max}$=9 min, 20 min, and 54 min for 16 nm, 25 nm, and 50 nm samples, respectively. The maxima of DRFs coincide with the As concentration maxima at the interface [Fig. 3 b)-d)]. The sputtering rate can be obtained by dividing $t_{max}$ by the thickness of overlayer. Estimated sputtering rates vary between 1.78 – 0.93 nm/min depending on the sample thickness.

## HAXPES of GaP(As)/n-Si(100) heterostructures

In Tab. SI-T5, HAXPES binding energies and VBOs measured on GaP(As)/n-Si(100) heterostructures are shown. HAXPES spectra are shown in Fig. 5 in the paper. Similar to GCIB-XPS measurements, the core level positions and VBOs do not depend strongly on GaP overlayer thickness. The VBM values, $E_{VBM}-E_F$, were measured on 50-nm-thick samples and n-Si(100):As references (0 nm). The VBM edges of the samples with thinner GaP overlayer were smeared by the emission from the Si substrate and, therefore, these values were excluded from the analysis. Averaged values of <VBO> were derived by using P 2p-Si 2p and Ga 3d-Si 2p core levels.

*Table SI-T4 - HAXPES binding energies measured on GaP(As)/n-Si(100) samples. The corresponding spectra are shown in Fig. 5 in the paper. Averaged values of valence band offsets, <VBO>, are shown in Fig. 6.*

| GaP thickness, nm | Substrate type | BE(P 2p$_{3/2}$), eV | BE(Ga 3d$_{5/2}$), eV | BE(Si 2p$_{3/2}$), eV | $E_{VBM}-E_F$, eV | VBO(P2p) | VBO(Ga3d) | <VBO>, eV | Source |
|---|---|---|---|---|---|---|---|---|---|
| 0 | | - | - | 99.85 | 1.0 | - | - | - | DESY |
| 4 | | 129.78 | 20.14 | 99.95 | - | 0.62 | 0.65 | 0.63 | SP-8 |
| 8 | | 129.78 | 20.13 | 99.94 | - | 0.63 | 0.65 | 0.64 | SP-8 |
| 16 | | 129.84 | 20.13 | 99.98 | - | 0.65 | 0.61 | 0.63 | SP-8 |
| 20 | n-type, 0.1°, A-type | 129.76 | 20.10 | 99.95 | - | 0.60 | 0.61 | 0.60 | SP-8 |
| 25 | | 129.77 | 20.12 | 99.96 | - | 0.60 | 0.62 | 0.61 | SP-8 |
| 50 | | 129.74 | 20.07 | - | 1.7 | - | - | - | SP-8 |
| 8 | | 129.75 | 20.10 | 99.94 | - | - | - | - | DESY |
| 50 | | 129.69 | 20.03 | - | 1.7 | - | - | - | DESY |

## HAXPES of GaP(As)/p-Si(100) heterostructures

HAXPES measurements were carried out on reference p-Si(100):As (0 nm) sample. Arsenic monolayers with A-type and B-type surface reconstructions were prepared on p-Si(100) deoxidized surfaces by MOVPE. Samples were transferred in air to DESY synchrotron (surface was oxidized). HAXPES measurements were carried out with an incidence energy of 6 keV. In Tab. SI-T5 and in Fig. SI-6, measured data are shown. The Si 2p core level positions of As-terminated surfaces are 99.4 - 99.5 eV and the VBM energy is 0.6-0.7 eV independently on type of As surface reconstruction.

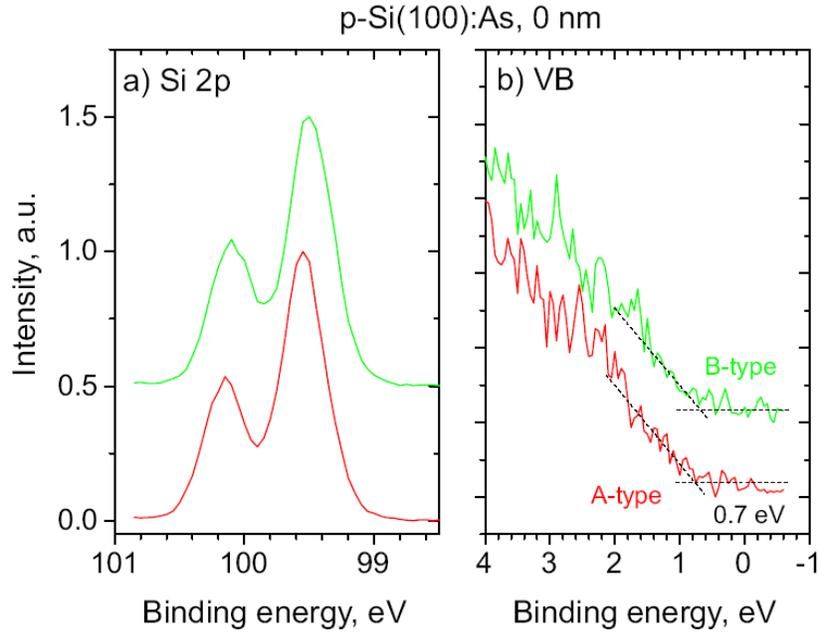

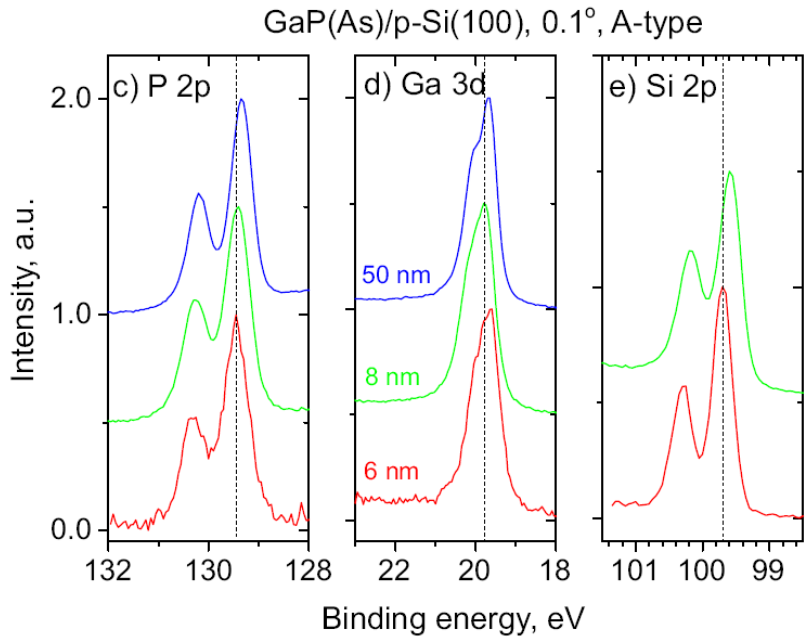

*Figure SI-6 – a) Si 2p core level spectra and b) VB spectra of 1 ML As layer on p-Si(100) substrate (0 nm samples) measured by HAXPES. c) P 2p, d) Ga 3d, and e) Si 2p core levels measured by HAXPES on GaP(As)/p-Si(100) samples.*

*Table SI-T5 - HAXPES binding energies measured on Si(100):As substrates and GaP(As)/p-Si(100) heterostructures. The corresponding spectra are shown in Fig. SI-6 and in Fig. SI-7.*

| GaP thickness, nm | Substrate type | BE(P $2p_{3/2}$), eV | BE(Ga $3d_{5/2}$), eV | BE(Si $2p_{3/2}$), eV | $E_{VBM}-E_F$, eV | VBO(P2p) | VBO(Ga3d) | <VBO>, eV | Source |
|---|---|---|---|---|---|---|---|---|---|
| 0 | p-Si(100):As, 0.1°, A-type | | | 99.37 | 0.6 | | | | DESY |
| 0 | p-Si(100):As, 4°, A-type | - | - | 99.54 | 0.7 | - | - | - | DESY |
| 0 | p-Si(100):As, 4°, B-type | - | - | 99.50 | 0.7 | - | - | - | DESY |
| 6 | p-type, 4°, B-type | 129.63 | 19.77 | 99.80 | - | 0.62 | 0.43 | 0.52 | DESY |
| 6 | p-type, 4°, A-type | 129.45 | 19.61 | 99.70 | - | 0.54 | 0.37 | 0.45 | DESY |
| 8 | p-type, 0.1°, A-type | 129.42 | 19.78 | 99.59 | - | 0.62 | 0.65 | 0.63 | DESY |
| 50 | p-type, 0.1°, A-type | 129.35 | 19.67 | - | 1.3 | - | - | - | DESY |

In Tab. SI-T5, HAXPES binding energies and VBO values measured on 6, 8, 50-nm-thick GaP(As)/p-Si(100) heterostructure samples are shown. The corresponding core level spectra are shown in Fig. SI-6 c)-d). There is weak variation of peak positions (~0.1 eV) toward low binding energy for thicker samples. Averaged value of 0.6 eV for 8 nm sample was obtained.

Fig. SI-7 a) shows valence band spectra (close to the VB edge) measured on p-Si(100):As (0 nm) sample and on 50-nm-thick GaP(As)/p-Si(100), 0.1°, A-type sample. The measured values are 1.3 eV for heterostructure and 0.6 eV for reference sample. This shift of 0.7 eV is close to the measured VBO of ~0.6 eV.

In Fig. SI-7 b), c), two band alignment models are shown. In b), downward band bending in Si substrate and upward band bending in GaP overlayer toward Si interface is present. Measured values of VBO and VBM are shown by red colour. Th VB edge of Si band at interface is thus at ~0.7 eV, which is close to the value measured on p-Si(100):As surface [green curve in a), 0.6 eV].

Core level shifts toward low binding energy were observed by HAXPES, however [Fig. SI-6 c)-e)]. This is an indication of upward (downward) band bending in substrate (overlayer) toward the interface. In Fig. SI-7 c), we suggest an alternative model for GaP(As)/p-Si(100) band alignment with local doping carrier variation at the interface: Arsenic atoms (n-type dopants) may interdiffuse into p-Si(100) substrate during MOVPE preparation and change charge carrier density locally in the interface proximity [6]. As result, local band bending variation is possible. Detailed analysis of band bending direction at such interface is still needed.

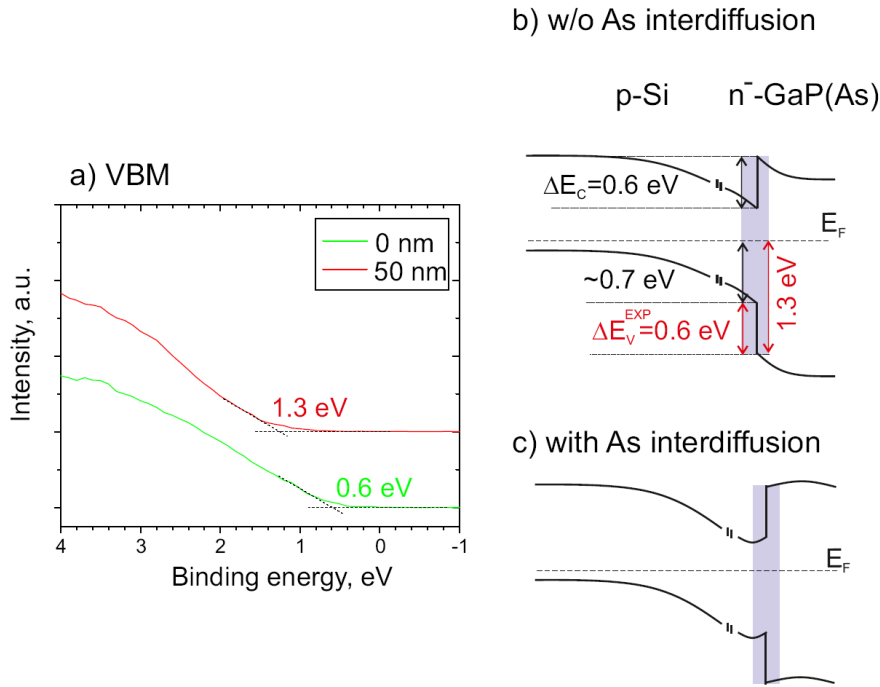

*Figure SI-7 – a) Valence band HAXPES spectra of the As:p-Si(100) (0 nm) and GaP(As)/p-Si(100) (50 nm) samples. Band alignment diagram models of b) without As interdiffusion and c) with As interdiffusion. Measured values of VBO and VBM are marked by red colour. A near interface region, which is assessable by the photoemission measurements, is schematically indicated by violet area.*

**Band gap states**

Figure SI-9 shows selected valence band spectra of the n-Si(100) substrate and the GaP(As)/n-Si(100) heterostructures. In b), a magnified region of the VB spectra is shown (~1% of normalized intensity). There is no signal close to the Fermi level (0 eV) for the bulk n-Si (black line) and 50-nm-thick (blues line) samples. Small increases in intensity were observed for 4 and 8 nm samples (indicated by red arrow), which could originate from the band-gap electronic states, which are localized at heterointerface and responsible for the pinning of the Fermi level.

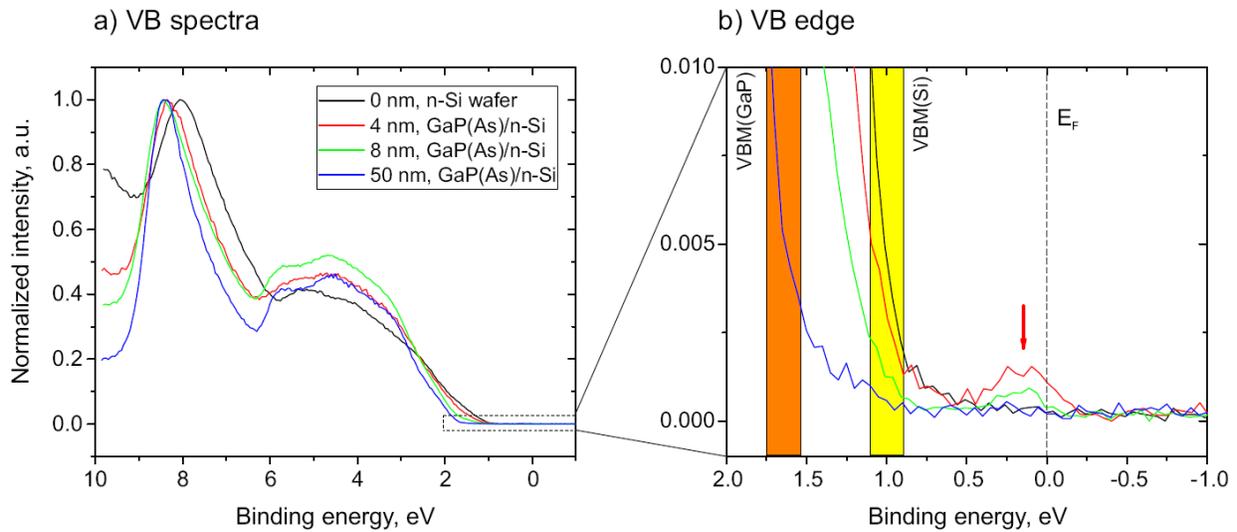

*Figure SI-9 – a) Normalized valence band spectra of n-Si(100) wafer and GaP(As)/n-Si(001) heterostructures. b) Magnified region of VB spectra close to VB edge. Fermi level and VBM of Si substrate and GaP overlayer are indicated. Arrow shows enhanced photoemission intensity close to the Fermi level.*